\documentstyle[prl,aps,preprint,epsf,floats]{revtex}
\begin{document}

\title{Josephson Plasma Resonance in High Temperature Superconductors: A Test of Interlayer
Tunneling Theory}

\author{Sudip Chakravarty}
\address{Department of Physics and Astronomy\\ University of California Los 
Angeles
\\ Los Angeles, CA 90095}
\date{\today}
\maketitle
\begin{abstract}
Josephson plasma frequency is computed within the 
interlayer tunneling theory and compared against the experimental results in
Tl$_2$Ba$_2$CuO$_y$ and La$_{2-{\rm x}}$Sr$_{\rm x}$CuO$_4$. It is shown that the
theoretical estimates are fully consistent with the recent experiments.
\end{abstract}
\pacs{PACS:}
\newpage
In the Josephson effect\cite{Josephson}, the quantum mechanical phases of two
superconductors are coupled by the tunneling of the Cooper pairs between them. The coupling
energy is a measure of the phase coherence, and the lowest energy is achieved when the phase
difference is zero. However, if a phase difference is imposed and the system is released,
the phase, to a good approximation, oscillates harmonically. Because the time derivative of
the phase is proportional to the voltage difference between the superconductors by the
Josephson relation, the voltage also oscillates harmonically, and the frequency of the
oscillation is called the Josephson plasma frequency, $\omega_J$\cite{Anderson1}. This is a
collective mode that has  the character of a longitudinal plasma wave. The frequency of this
wave is much smaller than the bulk  plasma frequency, which is due to the fact that the
junction region has a low density of charge carriers characteristic of the weak
superconductivity of a Josephson junction. Because $\omega_J$ is smaller than the bulk
plasma frequency, these oscillations are entirely confined within the junction region.
Such oscillations are well documented and were first measured by Dahm {\em et
al.}\cite{Dahm}. It has been shown\cite{Anderson} that the interlayer tunneling theory can
predict the frequency of the Josephson plasma frequency in high temperature
superconductors, where the plasma oscillations are confined between the two neighboring
layers. In addition, the magnitude of the plasma frequency contains important information
about the coupling between the layers, which is an important ingredient to the interlayer
tunneling theory of these superconductors\cite{Chakravarty}. This theory is applicable to
both $s$-wave and $d$-wave symmetries of the order parameter, as was stated in
Ref.~\cite{Chakravarty}; for explicit calculations involving $d$-symmetry, see
Refs.~\cite{Yin,Sudbo}.

In very recent optical experiments\cite{Marel}, it has been found that the Josephson 
plasma frequency in Tl$_2$Ba$_2$CuO$_y$ is smaller than 50 cm$^{-1}$. This is
much smaller than the earlier estimate 1500 cm$^{-1}$ based on the interlayer tunneling
theory\cite{Anderson}, which did not fully take into account the highly
anisotropic nature of the tunneling matrix element and the gap. When this
anisotropy is taken into account, I find that $\omega_J\le 282\ {\rm cm}^{-1}$.
The same calculation applied to La$_{2-{\rm x}}$Sr$_{\rm x}$CuO$_4$ leads to $\omega_J\le
80\ {\rm cm}^{-1}$  as compared to the measured value 50
${\rm cm}^{-1}$\cite{Marel,Uchida}. Thus,  given the uncertainty in the value of the
dielectric constant, $\epsilon$, I perceive no  difficulty with the interlayer tunneling
theory\cite{Leggett}. Note that because the Josephson plasma oscillations are confined
between the layers, the relevant dielectric constant is that of the material  between the
CuO planes. In addition, the present theory can be tested on other unconventional layered
superconductors, such as the layered organic superconductors\cite{KAM}, as well. It is also
worth noting that the conventional BCS dirty limit theory predicts $\omega_J$ to be 230
cm$^{-1}$ for Tl$_2$Ba$_2$CuO$_y$\cite{Marel}, while the same  dirty limit formula yields
200 cm$^{-1}$ for La$_{2-{\rm x}}$Sr$_{\rm x}$CuO$_4$.

The Josephson plasma frequency in the interlayer tunneling theory\cite{Chakravarty} 
is given by ($\bf k$ is the in-plane wavevector.)
\begin{equation}
{1\over 2}m\omega_J^2=\sum_{\bf k}T_J({\bf k})b_{\bf k}^2,
\end{equation}
where
\begin{eqnarray}
b_{\bf k}&=&{\Delta_{\bf k}\over 2E_{\bf k}}\tanh({E_{\bf k}\over 2 T}),\\
E_{\bf k}&=&\sqrt{(\varepsilon_{\bf k}-\mu)^2+|\Delta_{\bf k}|^2},\\
T_J({\bf k})&=&{T_J\over 16}(\cos k_xa-cosk_ya)^4,\\
m&=&{\hbar^2C\over 4e^2}.
\end{eqnarray}
The quantity $C$ is the capacitance of a parallel plate capacitor formed by the two
neigboring  layers of area $A$ and a separation $d$, that is,
\begin{equation}
C={\epsilon A\over 4 \pi d}.
\end{equation}
The quantity $T$ is the temperature, $\Delta_{\bf k}$ is the superconducting gap, 
$\varepsilon_{\bf k}$ is the single particle dispersion, and $\mu$ is the chemical
potential. Note that the change in the sign of the gap $\Delta_{\bf k}$ due to the
$d_{x^2-y^2}$ symmetry is unimportant in this formula.

Consider the $T\to 0$ limit. In the limit that the in-plane pairing is very small, a very
simple  expression for the magnitude of the $T=0$ gap is\cite{Yin}
\begin{equation}
|\Delta_k|=\sqrt{{T_J^2({\bf k})\over
4}-(\varepsilon_k-\mu)^2}\ \ \theta\left({T_J({\bf k})\over 2}-
|\varepsilon_{\bf k}-\mu|\right)\label{gap0}.
\end{equation} 
In the same limit, the transition temperature $T_c$ is given by $(T_J/ 4)$. When the
in-plane pairing is substantial, one must use the full gap equation; see Ref.~\cite{Yin}.

We get
\begin{equation}
(\hbar\omega_J)^2 =T_c{e^2 d\over 2\pi \epsilon a^2}I,
\end{equation}
where  $I$ is an integral that can be bounded by $I_0$, that is, $I\le I_0$, where 
\begin{eqnarray}
{I_0\over 4a^2}=\int_{0}^{\pi\over a}dk_x
\int_{0}^{{\pi\over a}}dk_y &&(\cos k_xa-\cos k_ya)^4\nonumber \\&&
\theta\left({T_J({\bf k})\over 2}- |\varepsilon_{\bf k}-\mu|\right),
\end{eqnarray}
Here $a$ is the lattice constant in the CuO-plane, and $\theta(x)=1$, for $x>0$, and equal
to 0 otherwise. Actually, $I_0$ should be  close to the true value because 
$|\varepsilon_{\bf k}-\mu|$ is very small in the region where the gap is finite. In any
case, by replacing $I$ by $I_0$, we are {\em overestimating} the plasma frequency.

I evaluate $I_0$\cite{I0} using  a typical band structure:
\begin{equation}
\varepsilon_{\bf k}=-2t\left[\cos (k_xa)+\cos (k_ya)\right]
+4t'\cos (k_xa)\cos (k_ya),
\end{equation}
with the 
parameters $t=0.25$ eV, $t'=0.45t$, and
$\mu=-0.315$ eV, corresponding to an open Fermi surface, with a
band filling of 0.86. The choice of these parameters is not critical to our
theory, nor do we believe that the van Hove singularity is a prominent
feature.

For Tl$_2$Ba$_2$CuO$_y$ , I choose  $T_c=(T_J/4)=85$ K and find that $I_0=2.4$. Then
\begin{equation}
(\hbar\omega_J)^2 \le 2.4T_c {de^2\over 2\pi \epsilon a^2}, 
\end{equation}
where $a=3.8${\AA}, $d=11.7${\AA}. It is
difficult to estimate precisely the relevant dielectric constant. However, if I choose
$\sqrt{\epsilon}\approx 5$, which is reasonable, I get
$\omega_J \le 282 {\rm cm}^{-1}$.

For La$_{2-{\rm x}}$Sr$_{\rm x}$CuO$_4$, $T_c=(T_J/4)=32 K$, which gives $I_0=0.88$. Then,
\begin{equation}
(\hbar\omega_J)^2 \le 0.88 T_c {de^2\over 2\pi \epsilon a^2}. 
\end{equation}
If I use $a=3.8${\AA}, $d=6.64${\AA}, and $\sqrt{\epsilon}\approx 5$, I get
$\omega_J \le 80 {\rm cm}^{-1}$.

The important point is that while $T_c$ is determined by $T_J/4$, which is half
the maximum $T=0$ gap, the plasma frequency depends on a momentum average of a highly 
anisotropic integrand which is nonvanishing only very close to the Fermi surface and
results in a low Josephson plasma frequency. 

I end with a few cautious remarks that even the
present estimates may not be fully accurate. First, and this is minor, I have used a
typical band structure. However, this cannot change the estimates much within the
interlayer tunneling theory because of the special shape of the gap. Second, the non-Fermi
liquid nature has not been fully taken into account except in so far as the single
particle tunneling is forbidden\cite{Chakravarty,Yin}. These effects are quite
complex in nature\cite{Chakravarty2} and are outside the scope of the present paper, but 
they can certainly make some difference. Third, the tunneling matrix element has been
assumed to be diagonal in the in-plane wavevector ${\bf k}$\cite{Chakravarty}.  For such a
$\bf k$-diagonal tunneling matrix element, the Josephson effect has a special discontinuous
feature when the gap tends to zero if calculated in the lowest order perturbation theory.
This feature can be corrected if one uses a renormalized perturbation theory such as the
Brillouin-Wigner perturbation theory\cite{Chakravarty3}. This should be of little
consequence for the present analysis, however. Finally, in keeping with the assumptions of
the interlayer tunneling theory, I have assumed that the in-plane pairing mechanism is a
small effect. However, even if we included this effect, as in Ref.~\cite{Yin}, the results
will not change much.


\begin{references}
\bibitem{I0}The integral $I_0$ was evaluated by a Monte Carlo method with $10^6$ sampling
points in the quarter of the Brillouin zone. This was necessary because the integrand is
very sharply peaked in a very narrow region of the Brillouin zone.
\end{references}
\end{document}